\documentclass[twocolumn,showpacs,preprintnumbers,amsmath,amssymb]{revtex4}

\usepackage{amsmath,amssymb} 
\usepackage[dvips]{graphicx} 
\usepackage[cp1251]{inputenc}

\newcommand\nn{\nonumber}
\newcommand\ba{\begin{eqnarray}}
\newcommand\ea{\end{eqnarray}}
\newcommand{\br}[1]{\left( #1 \right)}
\newcommand{\brs}[1]{\left[ #1 \right]}

\newcommand{\brm}[1]{\left| #1 \right|}

\newcommand{\MeV}{~\mbox{MeV}}

\begin{document}

\title{A model of a transition neutral pion formfactor measured in
annihilation and scattering channels}

\author{Yu.~M.~Bystritskiy}
\email{bystr@theor.jinr.ru}
\affiliation{JINR-BLTP, 141980 Dubna, Moscow region, Russian Federation}

\author{V.~V.~Bytev}
\email{bvv@jinr.ru}
\affiliation{JINR-BLTP, 141980 Dubna, Moscow region, Russian Federation}

\author{E.~A.~Kuraev}
\email{kuraev@theor.jinr.ru}
\affiliation{JINR-BLTP, 141980 Dubna, Moscow region, Russian Federation}

\author{A.~N.~Ilyichev}
\email{ily@hep.by}
\affiliation{National Scientific
and Educational Centre of Particle and High Energy Physics of the Belarusian State University,
220040  Minsk,  Belarus
}

\begin{abstract}
We consider an alternative explanation of newly found growth of neutral pion
transition form factor with virtuality of one of photon. It is based on Sudakov
suppression of quark-photon vertex.
Some applications to scattering and annihilation channels are considered including
the relevant experiments with lepton-proton scattering.
\end{abstract}


\maketitle

\section{Introduction}

A lot of attention was paid to the problem of describing the transition form factor
of neutral pion \cite{Chernyak:1977as,Lepage:1979zb,Efremov:1979qk}.
It is the information about wave function of neutral pion, namely the distribution
on the energy fractions of $u,d$ quarks inside a neutral pion, is the motivation of
numerous theoretical approaches to describe the transition form factor.
Recently some experimental information about it's behavior was obtained in process
$e^+e^-\to e^+e^-\pi_0$. The kinematics, when one of photon
is almost real and the other is highly virtual was considered \cite{Druzhinin:2009gq,:2009mc}.
The result was presented by authors of the experiment as a some nondecreasing function
of the module of square of momentum of a virtual photon. Such type of behavior is in
clear contradiction with the predictions of factorization theorem applying to this process
(see \cite{Mikhailov:2009kf,Radyushkin:2009zg,Polyakov:2009je} and references therein).

Below we consider another reason to explain such type of behavior, using the well known
expression of a virtual photon-quark vertex (so called Sudakov form factor \cite{Sudakov:1954sw} ) which is
entered in the triangle Feynman diagram, describing the conversion of two photon to the
neutral pseudoscalar meson. It is the motivation of this paper.

Both channels of pseudoscalar mesons production in elastic electron-positron
collisions - the scattering and annihilation ones are considered. The second one
$e^+e^-\to\pi_0 l^+l^-$ can be the subject of experimental investigation.

\begin{figure}
\scalebox{0.37}{\includegraphics{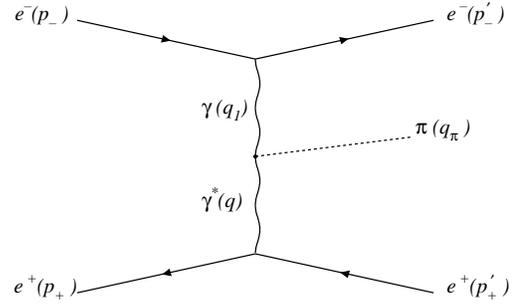}}
\caption{
Scattering process of pion production
}
\label{fig:sc}
\end{figure}
\section{Scattering channel}
\label{Scattering}

In the scattering type of experiments
\ba
    &&e^+\br{p_+} + e^-\br{p_-} \to
    e^+(p_+') e^-(p_-')\pi^0\br{q_\pi},
\label{sc}
\ea
($p_\pm^2=p_\pm'^{\; 2}=m_e^2$, $q_\pi^2=M^2$)
neutral pion is created
by two photons with momenta $q=p_+-p_+'$ and $q_1=p_--p_-'$ that
involving into lepton interaction as it is presented
on Fig.~\ref{fig:sc}. Due to Weizs\"acker-Williams (WW) kinematics
of this process (the scattered electron assumed to move close to beam direction)
one of photons is almost real
$\brm{q_1^2} \ll M^2$ and other is off mass shell $Q^2=-q^2 \gg M^2$.

Matrix element has a form (see for details in the Appendix~\ref{AppendixDetails})
\ba
M&=&\frac{2 s(4\pi\alpha)^2}{q^2q_1^2}\brs{\bf{q}\times\bf{q}_1}_z V(Q^2) N_+N_-, \nn\\
V(Q^2)&=&\frac{M_q^2}{2\pi^2 F_\pi Q^2}F\br{\frac{Q^2}{M_q^2}},
\label{VQ2}
\\
\brm{N_\pm}^2&=&\frac{1}{s^2}Tr\brs{p_-p_+p_-p_+} = 2,
\nn
\ea
where $s=(p_++p_-)^2 \gg Q^2$ and $V(Q^2)$ is the transition form-factor of pion,
$F_\pi=93\MeV$ is the decay constant of pion,
$M_q=M_u=M_d=280\MeV$ is the quark mass, $\bf{q}$, $\bf{q}_1$,
are transversal to the beam axes ($z$) direction components of
photon momenta.
\begin{figure}
\scalebox{0.37}{\includegraphics{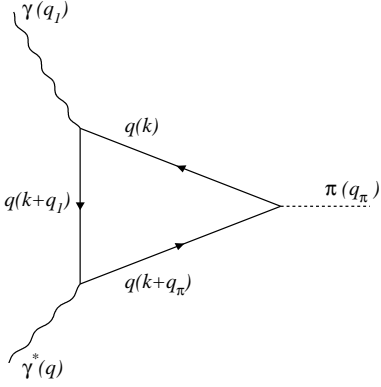}}
\caption{
Triangle vertex for $\gamma \gamma^* \to \pi _0$
process
}
\label{fig:ver}
\end{figure}
The quantity $F(Q^2/M_q^2)$ has a form:
\ba
&&F\br{Q^2/M_q^2}=-\int\frac{d^4 k}{i\pi^2}\times\nn\\
&&\times\frac{Q^2 V_S(Q^2, p_1^2,p_2^2)}
{(k^2-M_q^2+i0)(p_1^2-M_q^2+i0)(p_2^2-M_q^2+i0)},
\label{intdef}
\ea
where $p_1=k+q_1$, and $p_2=k+q_\pi$ and
Sudakov vertex function $V_S$ \cite{Carazzone:1975cc,Cornwall:1975ty} is:
\ba
V_S(Q^2, p_1^2,p_2^2)=
\exp\br{-\frac{\alpha_s C_F}{2\pi}\ln\frac{Q^2}{|p_1^2|}\ln\frac{Q^2}{|p_2^2|}},
\label{VSdef}
\ea
where $Q^2 \gg| p_{1,2}^2| \gg M_q^2$ and $C_F=\br{N^2-1}/\br{2N}=4/3$.
We use here the the Goldberger-Treiman relation on  the quark level
$F_\pi=M_q/g_{q\bar {q}\pi}=93\MeV$.

The cross section of the process (\ref{sc}) has a form:
\ba
d\sigma=\frac 1{8s}\brm{M}^2d\Gamma_3
\ea
The phase volume of the final state $d\Gamma_3$ can be expressed through
the Sudakov parametrization of the photon's momenta which turns out to be convenient:
\ba
q_1&=&\alpha_1 \tilde p_++\beta_1 \tilde p_-+q_{1\bot}, \nn\\
q&=&\alpha \tilde p_++\beta\tilde p_-+q_\bot, \label{SudPar}\\
a_\bot p_\pm&=&0, \quad
q_{1\bot}^2=-{\bf q}^2_{1\bot},\nn
\ea
and we imply the 4-vectors $\tilde p_\pm$ to be light-like, $2\tilde p_+\tilde p_-=s$.
Therefore
(details in Appendix~\ref{AppendixDetails}):
\ba
    d\Gamma_3
    &=&
    \br{2\pi}^{-5} \delta^4\br{p_+ + p_- - p_+' - p_-' - q_\pi}
    \times\nn\\
    &\times&
    \frac{d^3 p_+' d^3 p_-' q^3 q_\pi}{2 E_+' 2 E_-' 2 E_\pi}
    =
    \nn\\
    &=&
    \br{2\pi}^{-5} \frac{1}{4s} \frac{d\beta_1}{\beta_1\br{1-\beta_1}}
    d^2 q_1 d^2 q,
    \label{PhaseVolume}\\
    && \frac{Q^2+M^2}{s}<\beta_1<1. \nn
\ea
Using the expression for the square of momentum of "almost" real photon
\ba
q_1^2=-\frac{1}{1-\beta_1}\brs{{\bf q}^2_{1\bot}+m_e^2\beta_1^2},
\qquad
{\bf q}_{1\bot}^2 \ll Q^2,
\ea
and performing the integration on the parameters of scattered electron ($\beta_1,\vec{q}_1$),
moving close to $z$ axis we obtain for the cross section:
\ba
\frac{d\sigma}{d Q^2}&=&\frac{\alpha^4}{4Q^2}V^2(Q^2) J(Q^2), \nn \\
J(Q^2)&=&\frac{1}{2}L_s^2+L_s(L_e-1)-(L_e+1),
\ea
where
\ba
L_s=\ln\frac{s}{Q^2+M^2}, \qquad L_e=\ln\frac{Q^2}{m_e^2}.
\ea

There are several approaches to infer the value $V(Q^2)$, which is named
as pion transition formfactor.

One of them is based on QCD  collinear  factorization theorem \cite{Lepage:1979zb}
\ba
V^{BL}(Q^2)=\frac{2 F_\pi}{3}\int\limits_0^1\frac{dx}{x Q^2}\phi_\pi(x,s).
\ea
and in the papers  \cite{Mikhailov:2009kf, Polyakov:2009je}
different forms of pion wave function $\phi_\pi(x,s)$ was used.
Another possible mechanisms of the effect was given in \cite{Radyushkin:2009zg, Li:2009pr}.

Also in the paper \cite{Dorokhov:2009dg},\cite{Ametller:1983ec} was pointed
that pion form factor in the frames constituent quark model
has the double logarithmic asymptotic at large momentum transfer.

%
Another one which use the approach of Nambu--Jona-Lasinio model \cite{Volkov:1986zb,Volkov:2006vq}
gives:
\ba
V^{NJL}(Q^2)=\frac{2F_\pi}{3Q^2}.
\ea
The approach used here is based on the Sudakov form of the vertex function which describe
interaction of a photon with large four momentum square $|q^2|$ with two quarks of an anomalous
three angles quark diagram describing the conversion of two photons to the neutral pion.

The three angle  quark-loop diagramm itself at large $Q^2$ has the double-logarithm asymptotic  \cite{Ametller:1983ec} and
insertion of Sudakov form of the vertex function which includes also QCD-inspired
corrections gives  the asymptotic
$\ln\br{\ln\br{Q^2/m^2}}$
behavior.

\begin{figure}
\scalebox{0.9}{\includegraphics{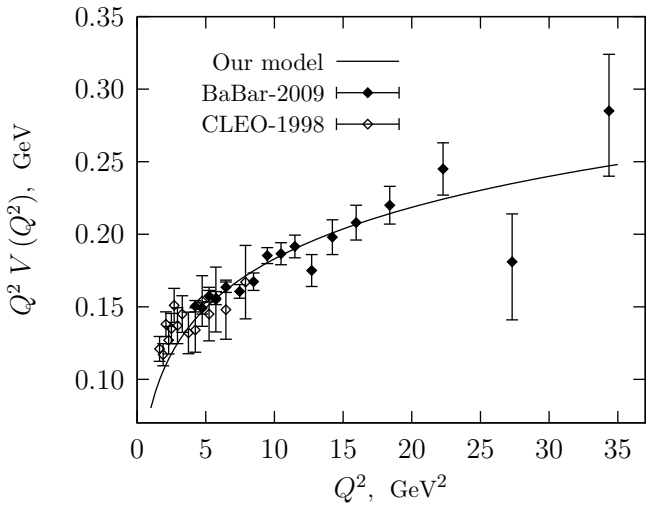}}
\caption{
A fit of our approach (Eq. (\ref{OurApproach})) with resultant fitting
parameters $A=0.49$, $B=0.23$ for $\pi^0$ production and the comparison with
the experimental data of BaBar \cite{:2009mc} and CLEO \cite{Gronberg:1997fj} facilities.
}
\label{fig:Fit}
\end{figure}

 The final expression is (details are in Appendix~\ref{AppendixDetails}):
\ba
V(Q^2)=A \frac{M_q^2}{2 \pi F_\pi \alpha_s C_F} \Phi(z_B),
\label{OurApproach}
\ea
where
\ba
\Phi(z)&=&\int\limits_0^1\frac{dx}{x}\br{1-e^{-z_B x(1-x)}}, \nn\\
z_B&=&\frac{C_F\alpha_s}{2\pi}\ln^2\frac{Q^2}{BM_q^2}, \nn
\ea
where $A$, $B$ can be considered as a positive fitting parameters of order of unity.
We find it through Babar data fitting.
Function $Q^2 V(Q^2)$ is presented in Fig. \ref{fig:Fit},
where the experimental data also presented.

\section{Annihilation channel}
\label{Annihilation}

Let us consider now the annihilation channel depicted on Fig.~\ref{fig:ann}:
\begin{figure}
\scalebox{0.4}{\includegraphics{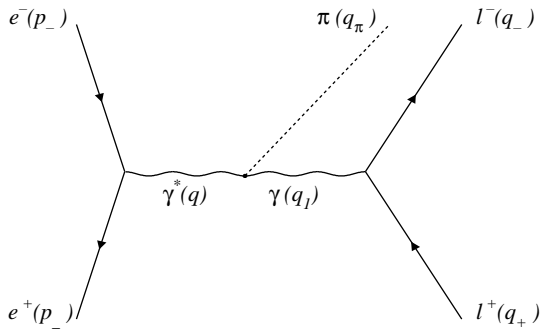}}
\caption{
The annihilation channel of process $e^+e^- \to \pi^0 l^+ l^-$.
}
\label{fig:ann}
\end{figure}
\ba
    e^+\br{p_+} + e^-\br{p_-} \to \gamma^*\br{q} \to \nn\\
    \to \pi^0\br{q_\pi} l^+\br{q_+}l^-\br{q_-},
\ea
where $l=e,\mu$ and $p_\pm^2=0$, $q_\pm^2=m_l^2$, $q_\pi^2=M^2$,
$s=q^2=(p_++p_-)^2$, $s_1=q_1^2=(q_++q_-)^2$.
We put the matrix element of this process in the form:
\ba
    M&=&\frac{\br{4\pi \alpha}^2}{q_1^2 q^2}
    J^\mu J^{\nu(l)} V\br{s}\epsilon_{\mu\nu\alpha\beta}q^\alpha q_1^\beta,
    \nn\\
    J^\mu&=& \bar v\br{p_+} \gamma_\mu u\br{p_-}, \nn\\
    J^{\nu(l)}&=& \bar v_\mu\br{q_+} \gamma_\nu u_\mu\br{q_-},
\ea
where quantity $V(s)$ describes conversion of two
off mass shell photons to the neutral pion (pion transition
formfactor) is defined in (\ref{VQ2}).

Let perform the phase volume of the final state
\ba
    d\Gamma_3 &=&
    \br{2\pi}^{-5} \delta^4\br{p_+ + p_- - q_+ - q_- - q_\pi}
    \times\nn\\
    &\times&
    \frac{d^3 q_+ d^3 q_- q^3 q_\pi}{2 E_+ 2 E_- 2 E_\pi},
\ea
 to the form
\ba
d\Gamma_3
    &=&
    \br{2\pi}^{-5} \pi d\Gamma_{q_1} dq_1^2 \frac{\Lambda^{1/2}\br{s,q_1^2,M_\pi^2}}{2s},
\ea
where $\Lambda(a,b,c)=a^2+b^2+c^2-2(ab+ac+bc)$ and
\ba
d\Gamma_{q_1} &=& \int \frac{d^3 q_+ d^3 q_-}{2 E_+ 2 E_-}
    \delta^4\br{q_1 - q_+ - q_-}.
\ea
Performing the integration on the invariant mass square of the lepton pair
(we use the approximation $s_1 \ll s$ and $s_1<M^2$)
the relevant cross section have a form (details in Appendix~\ref{AppendixDetails}):
\ba
\sigma^{e\bar{e}\to \pi_0 l\bar{l}}=
\frac{\pi\alpha^4 V(s)^2}{6}\br{1-\frac{M^2}{s}}^3
\brs{\ln\frac{s}{m_l^2}-\frac{5}{3}}.
\ea

\section{Conclusion}

From our point of view the QCD corrections connected with the vertex of interaction of
the highly virtual photon with quarks is essential and pion can be considered as a
point particle. So at rather large values of $Q^2$ the details of pion wave
function becomes irrelevant.

For heavy $s$ quarks $M_s=400\MeV$ entering the heavy pseudoscalar mesons $\eta'$
the effect of Sudakov form factor becomes more weak.

In literature presents the alternative explanations (see the end of
Section~\ref{Scattering})
of the experimental data BaBar \cite{:2009mc}.

On the Figure \ref{fig:Fit} we represent a numerical estimation fit of  the BaBar data.
We obtain the qualitative logaritm-logarithmical growth (see Eq. (\ref{asymptotics})) of the transitional form-factor (\ref{OurApproach}).
On the plot we put the best fitting of BaBar data with  two adjustable
parameters $A$ and $B$.

The similar phenomena can take place as well for the case of scalar mesons production.

We remind as well the possibility to measure the transition pion form factor
in electro-proton scattering $e p\to e\pi_0 p$. The relevant cross section will be
\ba
\frac{d\sigma^{ep\to e\pi_0 p}}{d Q^2}&=&
\br{\frac{\alpha g_{\rho qq}g_{\rho NN}}{8\pi(Q^2+M_\rho^2)}}^2
\frac{V^2(Q^2)}{Q^2}
\times\nn\\
&\times&\brs{F_1^2(Q^2)+\frac{Q^2}{4M_p^2}F_2^2(Q^2)}J(Q^2),
\ea
where $F_1$, $F_2$ -- are Dirac and Pauli proton form factors.
Here instead of virtual photon the virtual vector meson takes place; $g_{\rho qq}$,
$g_{\rho NN}$ are the $\rho$ meson couplings with quarks and nucleons correspondingly.
In this case a problem with background
($ep\to e\Delta^+\to e\pi^0 p$)
must be overcomed.

We consider Sudakov form factor for time-like transfer momentum. With
ordinary particles in the loop we must take into account the imaginary part of relevant amplitude.
For quarks inside a loop imaginary part is absent.

Taking into account the non-leading terms in expression of Sudakov exponent results in
modification of quark mass and a general shift of normalization:
\ba
F\br{q^2,p_1^2,p_2^2} &\to& A F\br{q^2,p_1^2,p_2^2},\\
Q^2\alpha\beta &>& B M_Q^2, \nn
\ea
with $A\sim B\sim 1$ can be considered as a fitting parameters $A,B>0$.

\begin{acknowledgments}
 We thank Dr. S.~V.~Mikhailov for his interest to the subject.
\end{acknowledgments}

\appendix

\section{Details of calculation}
\label{AppendixDetails}

Transformation of the phase volume of 2-gamma creation process $2\to 3$ consist
in introduction of two transferred vectors $q_1,q$:
\ba
d\Gamma_3 &=& (2\pi)^{-5}d^4q_1d^4qd^4q_\pi d^4p'_+d^4p'_-\times \nn \\
&\times& \delta^4\br{p_--q_1-p'_-} \delta^4\br{p_+-q-p'_+}
\times \nn\\
&\times& \delta^4\br{q+q_1-q_\pi} \delta\br{\br{q_1-p_-}^2-m_e^2}
\times\nn\\
&\times&\delta\br{\br{q-p_+}^2-m_e^2}\delta\br{\br{q_1+q}^2-M^2},
\nn
\ea
 and using the Sudakov parametrization
for the scattering channel (\ref{SudPar})
 we put it in form:
\ba
d\Gamma_3 &=& (2\pi)^{-5}\frac{s}{2}d\alpha_1 d\beta_1
d^2\vec{q}_1\frac{s}{2}d\alpha d\beta d^2\vec{q}
\times\nn\\
&\times&
\delta\br{s\alpha\beta-\vec{q}^2-s\beta-m_e^2\alpha}
\times\nn \\
&\times&
\delta\br{s\alpha_1\beta_1-\vec{q}_1^2-s\alpha_1-m_e^2\beta_1}
\times\nn\\
&\times&
\delta\br{s\alpha\beta_1-\br{\vec{q}+\vec{q}_1}^2-M^2}.
\ea
Performing the integrations over $\alpha_1$, $\alpha$, $\beta$,
we obtain the result given above
(see Eq.~(\ref{PhaseVolume})).

Expression for scalar loop integral with 3 denominators and Sudakov vertex
inserted has a form:
\ba
    F\br{\frac{Q^2}{M_q^2}}&=&-\int \frac{d^4 k}{i\pi^2}
    \frac{Q^2 V_S(Q^2,p_1^2,p_2^2)}{(k)(1)(2)}, \nn\\
    (k)&=&k^2-M_q^2+i0; \nn\\
    (1)&=&(k+q_1)^2-M_q^2+i0; \nn\\
    (2)&=&(k+q_\pi)^2-M_q^2+i0. \nn
\ea
where $V_S(Q^2,p_1^2,p_2^2)$ was defined in (\ref{VSdef}).

To perform the integration we use Sudakov parametrization of the loop momentum:
\ba
k=\alpha n_1+\beta n_2+k_\bot,
\ea
with $n_{1,2}$ are light-like 4 vectors $n_i^2=0$, transversal to $k_\bot$, $ n_i k_\bot=0$,
builded from the 4-vectors $q_\pi$, $q_1$ such that $2n_1n_2=Q^2$.
In such a parameterization
\ba
d^4k=\frac{Q^2}{2}d\alpha d\beta d^2k_\bot.
\ea
Expressing the denominators of quark Green functions:
\ba
(k)=Q^2\alpha\beta-\vec{k}^2-M_q^2+i0; \quad (1)\approx Q^2\alpha; \quad (2)\approx Q^2\beta, \nn
\ea
and performing the integration over $k_\bot^2=-\vec{k}^2$ as (we imply that principal part
does no contribute)
\ba
&&\int\frac{\pi d\vec{k}^2}{(k)}=-i\pi^2 \int d\vec{k}^2 \delta(-\vec{k}^2+Q^2\alpha\beta-M_q^2)=
\nn\\
&&=
-i\pi^2\theta[Q^2\alpha\beta-M_q^2],
\ea
we obtain
\ba
F\br{\frac{Q^2}{M_q^2}}&=&\frac{1}{2}\int\limits_{M_q^2/Q^2}^1\frac{d\alpha}{\alpha}\int\limits_{M_q^2/Q^2}^1\frac{d\beta}{\beta}
~\theta\br{Q^2\alpha\beta-M_q^2}\times\nn\\
&\times&
\exp\br{-\frac{\alpha_s C_F}{2\pi}\ln\frac{1}{\alpha}\ln\frac{1}{\beta}}.
\ea
Performing one integration we obtain:
\ba
F(\frac{Q^2}{M_q^2})=\frac{\pi}{\alpha_s C_F}\Phi(z),
\ea
where
\ba
&&\Phi(z)=\int\limits_0^1\frac{d x}{x}\br{1-e^{-zx(1-x)}}, \nn\\
&&z=\frac{\alpha_s C_F}{2\pi}L^2, \quad
L=\ln\frac{Q^2}{M_q^2}. \nn
\ea
For large values of $Q^2$ we obtain
\ba
F(\frac{Q^2}{M_q^2})=\frac{\pi}{\alpha_s C_F}\br{\ln\br{z}+c},
\label{asymptotics}
\ea
where
\ba
c=\int\limits_0^1\frac{d x}{x}(1-e^{-x})-\int\limits_1^\infty
\frac{d x}{x e^x}\approx 0.57.
\ea
In reality the quantity $L\sim 5$ for light consistent quarks $u,d$ which are present in the
neutral pion and  $L\sim 1-2$ for $s$-quark in $\eta'$ meson.

When considering the integration on the pair phase volume in annihilation
channel we use the relation (consequence of gauge invariance):
\ba
&&\sum_{pol}\int d\Gamma_{q1}J^{(\mu (l))}(J^{(\nu (l))})^*=
\nn\\
&&=-\frac{1}{3}\br{g_{\mu\nu}-\frac{q_{1\mu}q_{1\nu}}{q_1^2}}\br{q_1^2+2m_l^2}
\frac{\pi\beta_-}{2}, \\
&&
\beta_-=\sqrt{1-\frac{4m_l^2}{q_1^2}}. \nn
\ea
The differential cross section is
\ba
d\sigma^{e\bar e\to \pi^0 l \bar l}&=&\frac{\alpha^4M_q^4}{24\pi^3F_\pi^2s^5}
\Lambda^{3/2}\br{s,q_1^2,M^2}\frac{d x}{x}\times\nn\\
&\times&
\br{1+\frac{1}{2x}}
\sqrt{1-\frac{1}{x}},
\ea
where $x=q_1^2/\br{4m_l^2}>1$.


\end{document}